\documentclass[aps,prl,reprint,superscriptaddress, showpacs]{revtex4-1}

\usepackage{textcomp}
\usepackage{amsmath}
\usepackage{amssymb}
\usepackage{mathrsfs}
\usepackage{color, soul}
\usepackage{empheq}
 
\usepackage{graphicx}

\begin{document}

\title{Reinventing atomistic magnetic simulations with spin-orbit coupling}

\author{Dilina Perera}
\email[]{dilinanp@physast.uga.edu}
\affiliation{Center for Simulational Physics, The University of Georgia, GA 30602, USA}

\author{Markus Eisenbach}
\affiliation{Oak Ridge National Laboratory, Oak Ridge, TN 37831, USA}

\author{Don M. Nicholson}
\affiliation{University of North Carolina at Asheville, Asheville, NC 28804, USA}

\author{G. Malcolm Stocks}
\affiliation{Oak Ridge National Laboratory, Oak Ridge, TN 37831, USA}

\author{David P. Landau}
\affiliation{Center for Simulational Physics, The University of Georgia, GA 30602, USA}


\begin{abstract}
We propose a powerful extension to combined molecular and spin dynamics that fully captures the coupling between the atomic and spin subsystems via spin-orbit interactions. 
Its foundation is the inclusion of the local magnetic anisotropies that arise as a consequence of the lattice symmetry breaking due to phonons or defects. 
We demonstrate that our extension enables the exchange of angular momentum between the atomic and spin subsystems, 
which is critical to the challenges arising in the study of fluctuations and non-equilibrium processes in complex, natural, and engineered magnetic materials.
\end{abstract}


\maketitle

With the prevailing computational cost of first-principles based methods, 
there is a continued demand for atomistic simulations
as a viable approach for predicting the finite-temperature properties of materials.
To this end, the molecular dynamics (MD) method~\cite{Rapaport2004, Frenkel2002} has long been the de-facto standard
for modeling the time evolution of material structure,
providing quantitative insight into a wide range of physical phenomena, 
including radiation damage cascades~\cite{RadiationCascades1, RadiationCascades2}, fracture behavior~\cite{Fracture1}, 
dislocation dynamics~\cite{DislocationDynamics1}, self diffusion~\cite{SelfDiffusion1} etc.
The lesser-known counterpart for probing magnetic properties is the spin dynamics (SD) method~\cite{Gerling1990, Chen1994, Landau1999, Zhuofei2015},
in which one models the magnetic crystal as a classical system of interacting atomic magnetic moments on a rigid lattice.
With parameterized exchange coupling,
SD renders a powerful means of characterizing collective magnetic excitations
with good quantitative agreement with the experiments~\cite{Tao2005, Tsai2000}.

However, due to strong spin-lattice coupling observed in transition magnetic metals and alloys~\cite{SpinLatticeCoupling1, SpinLatticeCoupling2}, 
the validity of MD and SD as stand-alone simulation methods is highly debatable.
For instance, in Iron-based materials, phonon-magnon coupling plays a pivotal role in maintaining the structural stability~\cite{PhaseStabilityIron, PhaseStabilityFeCo}, 
and significantly influences the thermal transport properties~\cite{ThermalTransport}, defect evolution~\cite{DefectEvolution},
and the equilibrium thermodynamic behavior~\cite{Thermodynamics}. 
Thus, for a realistic depiction of a magnetic crystal, it is imperative that the dynamics of translational and spin degrees of freedom are treated on an equal footing.
The recently introduced ``spin-lattice dynamics'' or ``combined molecular and spin dynamics (MD-SD)'' approach~\cite{Ma2008}
establishes a robust computational framework for the aforementioned unification of MD and SD. 
The method has been successfully applied for bcc Iron with emphasis on
phonon-magnon interactions~\cite{Perera2014, PereraST2014},
vacancy formation and migration~\cite{WenVacancy1, WenVacancy2},
and external magnetic field effects~\cite{MagField}.
Moreover, an adaptation of MD-SD has been recently applied to cobalt nanosystems with large shape anisotropies~\cite{AnisCobalt}.

Despite wide applicability, the MD-SD formalism suffers from a fundamental flaw  
that prohibits angular momentum exchange between the lattice and the spin subsystems~\cite{Ma2008}. 
This inhibits the modeling of the spin-lattice relaxation process, with profound implications in non-equilibrium simulations.
In this paper, we discuss this aspect in detail and present an extension to MD-SD that eliminates this problem. 
The proposed solution relies on a valuable concept that is absent in the traditional approach: the introduction of a local anisotropy term 
to capture the effect of the spin-orbit interaction due to the symmetry breaking of the local atomic environment.

In the conventional approach to MD-SD, the material is modeled as a classical system of $N$ magnetic atoms of mass $m$, described by the Hamiltonian 
\begin{equation} \label{eq:hamiltonian}
  \mathcal{H} = \sum_{i=1}^N \frac{ m{\mathrm{v}_i}^2 }{2} + U(\{\mathbf{r}_i\}) - \sum_{i<j} J_{ij}(\left\{ \mathbf{r}_k \right\}) \mathbf{S}_i \cdot \mathbf{S}_j,
\end{equation}
where $\{\mathbf{r}_i\}$, $\{\mathbf{v}_i\}$ and $\{\mathbf{S}_i\}$ are the  positions, velocities and classical spins, respectively.
$U(\{\mathbf{r}_i\})$ is the non-magnetic component of the interatomic potential
whereas the Heisenberg-like interaction with the coordinate-dependent exchange parameter $J_{ij}(\left\{ \mathbf{r}_k \right\})$
specifies the exchange coupling between the spins.

The time evolution of the phase variables is governed by the coupled equations of motion
\begin{subequations}
\label{eq:eom}
\begin{empheq}[left=\empheqlbrace]{align}
\frac{d\mathbf{r}_i}{dt} &= \mathbf{v}_i 							\label{eq:eom_pos}          	   	  \\[6pt]
\frac{d\mathbf{v}_i}{dt} &= \frac{\mathbf{f}_i}{m}						\label{eq:eom_vel} \qquad \qquad \quad , \\[6pt]
\frac{d\mathbf{S}_i}{dt} &= \frac{1}{\hbar{}} \mathbf{H}_i^\text{eff} \times \mathbf{S}_i	\label{eq:eom_spin}
\end{empheq}
\end{subequations}
where $\mathbf{f}_i = -\nabla{}_{\mathbf{r}_i} \mathcal{H}$ and $\mathbf{H}_i^\text{eff} =\nabla{}_{\mathbf{S}_i} \mathcal{H}$ 
are the interatomic force and the effective field, respectively.
In MD-SD, one seeks to numerically solve these equations and obtain the trajectories of the atomic and spin degrees of freedom.
With $U(\{\mathbf{r}_i\})$ and  $J_{ij}(\left\{ \mathbf{r}_k \right\})$ chosen appropriately, one can readily adopt this model to
any magnetic material in which the spin interactions can be modeled classically.
For demonstration purposes, we will use the adaptation of Ma \textit{et al.}~\cite{Ma2008} for bcc iron, 
in which $U(\{\mathbf{r}_i\})$ is constructed as
$U(\{\mathbf{r}_i\}) = U_{\text{DD}} - E_{\text{spin}}^{\text{ground}},$
where $U_{\text{DD}}$ is the Dudarev-Derlet embedded atom potential~\cite{Dudarev2005, Derlet2007},
and $E_{\text{spin}}^{\text{ground}} = - \sum_{i<j} J'_{ij}(\left\{ \mathbf{r}_k \right\})$ is the energy contribution from a collinear spin state
which avoids the double counting of the spin-spin interaction,
with $J'_{ij}(\left\{ \mathbf{r}_k \right\}) = J_{ij}(\left\{ \mathbf{r}_k \right\})|\mathbf{S}_i||\mathbf{S}_j|$ being
the modified exchange interaction with the spin lengths absorbed into its definition.
For $J'_{ij}(\left\{ \mathbf{r}_k \right\})$, we use a pairwise functional form $J'(r_{ij})$ 
parameterized by first principles calculations~\cite{Ma2008}.
For simplicity, we assume constant spin lengths $|\mathbf{S}| = 2.2/g$, where $g$ is the electron $g$ factor. 

In MD-SD, the coupling between the lattice and the spin subsystem is established via
the coordinate-dependence of the exchange interaction, 
which allows the exchange of energy between the two subsystems.
However, this exchange coupling alone does not facilitate the transfer of angular momentum.
Due to the rotational symmetry of the Hamiltonian, in the absence of any external torques that explicitly perturb the spin orientations, 
the total spin angular momentum remains a constant of motion, irrespective of the dynamics of the lattice subsystem.
In non-equilibrium simulations, this unrealistic constraint may impose an entropic barrier between the two subsystems
and prevent them from achieving the mutual equilibrium.

\begin{figure}[ht]
 \begin{center}
  \includegraphics[width=0.85\columnwidth]{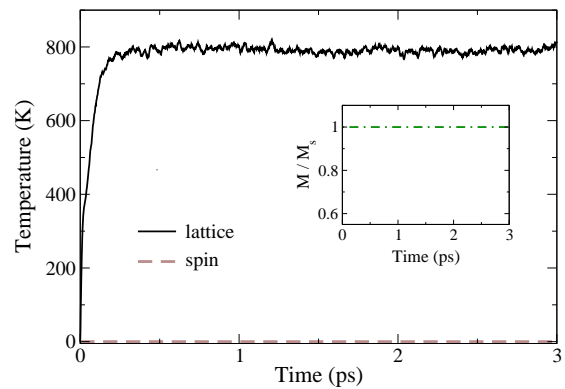}
 \end{center}
  \caption{
	(Color online) Thermal relaxation in a conventional molecular and spin dynamics simulation,
	with the lattice subsystem coupled to a heat bath at the temperature $T = 800$\;K. 
	The instantaneous lattice and the spin temperatures are plotted as functions of time.	
	The inset shows the time evolution of magnetization as a fraction of the saturation magnetization.	
	} 
 \label{fig:failed_thermalization}
\end{figure}

To demonstrate this point, we investigate the thermalization of a coupled spin-lattice system via an external heat bath
that interacts exclusively with the lattice subsystem.
If spin-lattice coupling is properly established, 
a heat bath connected to either of the subsystems should allow them both to thermalize towards the same equilibrium temperature. 
Dimensions of the simulation cell were chosen to be $16 \times 16 \times 16$, with periodic boundary conditions along $x$, $y$ and $z$ directions.
Initially, all atoms were arranged on a perfect bcc lattice with spins oriented along the $z$ direction, and velocities set to zero.
The heat bath was modeled using the stochastic Langevin dynamics equation for the translational degrees of freedom~\cite{Kubo1966}.
Coupled equations of motion in Eq.~\eqref{eq:eom} were integrated by an algorithm based on 
the second order Suzuki-Trotter (ST) decomposition of the non-commuting operators \cite{Omelyan2001, Tsai2005, Krech1998}, 
using a time step of $\delta t = 1$\,fs.
Fig.~\ref{fig:failed_thermalization} shows the time evolution of the instantaneous temperatures
associated with the lattice and the spin subsystems, with the target temperature of the thermostat set to $800$\,K.
Spin temperature was measured using the formula developed by Nurdin \textit{et al.}~ \cite{Nurdin2000}.
Due to the direct contact with the heat bath, the lattice subsystem thermalizes and reaches the equilibrium within a fraction of a picosecond.
However, the currently established form of spin-lattice coupling fails to initiate the thermal excitation of the spin orientations,
constraining the spin temperature and the magnetization (shown in the inset) to remain constant throughout the simulation.

The above example highlights the need for exploring missing contributions to spin-lattice coupling
that may potentially capture the true dynamics of the relaxation process.
An important interaction currently excluded from MD-SD as presented in Eq.~\eqref{eq:hamiltonian} is spin-orbit (SO) coupling,
which serves as a direct channel for the flow of energy and angular momentum between the spins and the lattice~\cite{Stohr2006}. 
However, in the bulk phase of $3d$ cubic transition metals and alloys,
crystal-field splitting largely suppresses SO interactions~\cite{Skomski2008, Mohn2003}, 
leading to coupling strengths that are several orders of magnitude smaller than the exchange interaction~\cite{Stohr2006}.
Nevertheless, in low symmetry environments such as surfaces and thin films,
SO coupling is considerably strengthened 
due to the the changes in the periodic potential experienced by the electrons~\cite{Wang1993, Lessard1997}.
Following the same argument, we assert that the momentary symmetry breaking of the crystal structure
that occurs due to phonons may substantially enhance SO interactions.
This, in turn, may significantly influence the spin-lattice relaxation process at elevated temperatures.

We infer that the critical ``missing piece'' in MD-SD is in fact a classical model that encapsulates
such local fluctuations in the SO interactions.
As the MD-SD formalism does not contain the notion of orbital angular momentum ($\mathbf{L}$),
one cannot introduce SO coupling directly in its natural form, $\mathcal{H}_{\text{SO}} \sim \mathbf{L} \cdot \mathbf{S}$.
Therefore, we model the effect of SO coupling via one of its emergent properties, magnetocrystalline anisotropy~\cite{Daalderop1990,WangAnis1993}.
As the effective size and the orientational preference of the SO interaction depends on the symmetry of the surrounding atomic environment,
the resultant anisotropies will also vary across different atomic sites.
Magnitudes and the easy axes of these ``induced'' local anisotropies will change dynamically
as the local environment is continuously distorted by the propagating phonons.

Henceforth, we will neglect any ``background'' anisotropy that already reside in the perfect crystalline symmetry,
and only focus on the aforementioned induced anisotropies.
Based on the first and the second order terms of the anisotropy energy expansion for a single spin,
we propose the following terms for the anisotropic components of the Hamiltonian
\begin{equation}\label{eq:anis_hamiltonian}
	\mathcal{H}_{\text{anis}} = - C_1 \sum_{i=1}^N \mathbf{K}_i \cdot \mathbf{S}_i 
				- C_2 \sum_{i=1}^N \mathbf{S}_i^\intercal \cdot \mathbf{\Lambda}_i \cdot \mathbf{S}_i ,
\end{equation}
where $C_1$ and $C_2$ are constants,
and vector $\mathbf{K}_i$ and tensor $\mathbf{\Lambda}_i$ are variable quantities 
that define the easy axes and the coupling strengths of the on-site magnetic anisotropy at a given time.
$\mathbf{K}_i$ and $\mathbf{\Lambda}_i$ are solely determined by the symmetry of the local atomic environment.
Since we have chosen to ignore the cubic anisotropy present in the ground state,
these quantities will vanish for perfect cubic crystalline symmetry.
In order to establish the connection to the local environment, we write the vector $\mathbf{K}_i$ and tensor $\mathbf{\Lambda}_i$ as
\begin{equation}
		\mathbf{K}_i = \nabla_{\mathbf{r}_i} \rho_i \; \; ,  \qquad	
		\mathbf{\Lambda}_i = \begin{pmatrix}
			\frac{\partial^2\rho_i}{\partial {x_i}^2} & \frac{\partial^2\rho_i}{ \partial {x_i} \partial {y_i} } 
						& \frac{\partial^2\rho_i}{ \partial {x_i} \partial {z_i} } \\[0.5em]
			\frac{\partial^2\rho_i}{ \partial {y_i} \partial {x_i} } & \frac{\partial^2\rho_i}{\partial {y_i}^2} 
					  	& \frac{\partial^2\rho_i}{ \partial {y_i} \partial {z_i} } \\[0.5em]
			\frac{\partial^2\rho_i}{ \partial {z_i} \partial {x_i} } & \frac{\partial^2\rho_i}{ \partial {z_i} \partial {y_i} }
					  	& \frac{\partial^2\rho_i}{\partial {z_i}^2} 
			  	\end{pmatrix} ,
\end{equation}
where $\rho_i(\{\mathbf{r}_k\})$ is a scalar function that quantitatively reflects the local symmetry surrounding the $i$th atom.
The particular functional form of $\rho_i(\{\mathbf{r}_k\})$  will depend on the details of the electronic structure of the material.
Since the first principles based formulation of $\rho_i(\{\mathbf{r}_k\})$ is challenging the capabilities of the current \textit{ab initio} methods,
we phenomenologically construct $\rho_i(\{\mathbf{r}_k\})$ as
$\rho_i = \sum_{j (j \neq i)} \phi(r_{ij})$, where $\phi (r_{ij})$ is an arbitrary pairwise function.
The chosen functional form assures that in perfect cubic crystalline symmetry, $\nabla_{\mathbf{r}_i} \rho_i$ 
and the off-diagonal elements of $\mathbf{\Lambda}_i$ vanish.
While the diagonal elements of $\mathbf{\Lambda}_i$ do not vanish, they become identical, 
which only contributes to a constant shift in the ground state energy.
For $\phi (r_{ij})$, we choose a short-range function
\begin{equation}
	\phi (r_{ij}) = 
			\begin{cases}
    				(1-r_{ij}/r_c)^4 \exp(1-r_{ij}/r_c), & \text{$r_{ij} \leq r_c$}.\\
    				0, & \text{$r_{ij}>r_c$}.
  			\end{cases} ,
\end{equation}
with the cut-off distance $r_c = 3.5$\;\AA\, between the second and the third nearest neighbor distances of the bcc iron lattice.
The fourth-order polynomial component ensures that all interatomic forces
due to the coordinate-dependence of $\mathcal{H}_{\text{anis}}$ smoothly approach zero at $r_c$.

First principles methods such as Locally Self Consistent Multiple Scattering (LSMS)~\cite{lsms} can routinely provide SO energies
associated with the vibrational breaking of local symmetry, and hence; in principle; estimates for the coefficients $C_1$ and $C_2$.
An attempt at parameterizing $C_1$ based on LSMS calculations~\cite{JunqiUnpublished} of a $128$ atom configuration with thermal displacements 
yielded an average value in the order of $10^{-1}$\;eV, with a site-to-site root-mean-square deviation of the same order.
Such variation from site-to-site demonstrates the difficulty in extracting models for SO energies 
from the overall energy shifts associated with the local displacements as predicted by LSMS.
Therefore, in what follows, we choose values for $C_1$ and $C_2$ of the order of $10^{-1}$\;eV,
and further explore the sensitivity of the results to their variations.

Eq.~\eqref{eq:hamiltonian} combined with the anisotropy terms in Eq.~\eqref{eq:anis_hamiltonian}
establishes a complete MD-SD model that fully couples the atomic and spin degrees of freedom.
The proposed extension preserves the conservation laws of the original model, 
including the conservation of energy, linear momentum, and total angular momentum.
With the inclusion of the second-order anisotropy term, Eq.~\eqref{eq:eom_spin} becomes non-linear, 
rendering the conventional ST algorithm inapplicable.
To circumvent this issue, we use a hybrid integration method that combines the  ST decomposition 
with the iterative scheme proposed by Krech et al.\ \cite{Krech1998}.  
To obtain the same level of accuracy as reflected by the energy conservation in microcanonical simulations,
we reduce the integration time step to $\delta t = 0.1$\,fs. 

\begin{figure}[ht]
 \begin{center}
  \includegraphics[width=0.85\columnwidth]{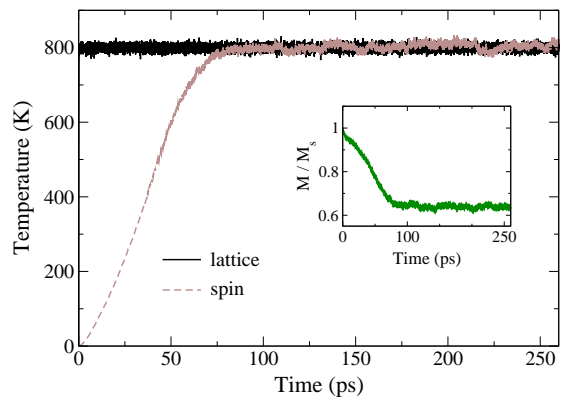}
 \end{center}
  \caption{
	(Color online) Thermal relaxation in a combined molecular and spin dynamics simulation enhanced with SO coupling.
	The lattice subsystem is coupled to a heat bath at the temperature $T = 800$\;K.
	Anisotropy coefficients $C_1$ and $C_2$ were set to $0.2$\;eV and $0.1$\;eV, respectively.
	} 
 \label{fig:anis_thermo_test}
\end{figure}

\begin{figure}[ht]
 \begin{center}
  \includegraphics[width=0.85\columnwidth]{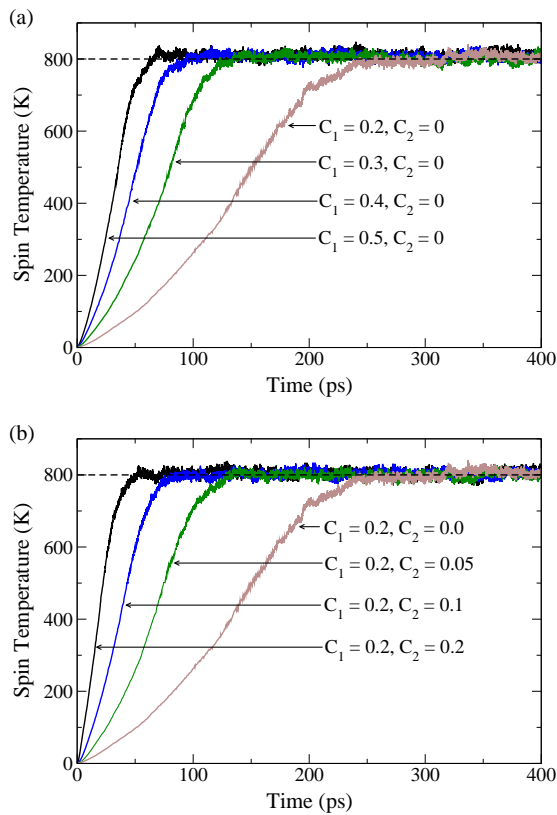}
 \end{center}
  \caption{
	(Color online) Thermalization of the spin subsystem under varying anisotropy strengths:
	(a) varying the first order anisotropy coefficient $C_1$ while the second order coefficient $C_2$ set to zero,
	(b) varying $C_2$ while $C_1$ held constant at $0.2$\;eV. 
	The lattice subsystem is coupled to a heat bath at the temperature $T = 800$\;K.
	} 
 \label{fig:anis_strength}
\end{figure}

\begin{figure}[ht]
 \begin{center}
  \includegraphics[width=0.85\columnwidth]{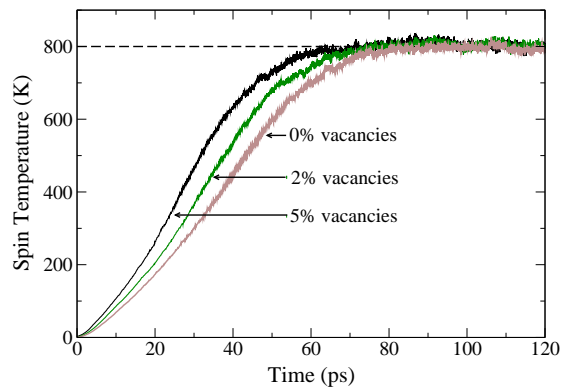}
 \end{center}
  \caption{
	(Color online) Thermalization of the spin subsystem under varying vacancy concentrations,
	with lattice subsystem coupled to a heat bath at the temperature $T = 800$\;K.
	Anisotropy coefficients $C_1$ and $C_2$ were set to $0.2$\;eV and $0.1$\;eV, respectively.
	} 
 \label{fig:vacancies}
\end{figure}

To show that our novel extension eradicates the barrier for the angular momentum exchange, 
we perform the previously described thermalization procedure,
with the anisotropy coefficients $C_1$ and $C_2$ set to $0.2$\;eV and $0.1$\;eV, respectively.
The corresponding results are shown in Fig.~\ref{fig:anis_thermo_test}.
As anticipated, the spin subsystem gradually loses angular momentum to the lattice through the anisotropy terms,
allowing the spin temperature/magnetization to increase/decrease with time.
The precessional damping of the spins continues until the coupled spin-lattice system; as a whole;
approaches equilibrium with the phonon heat bath.

With the pairwise function $\phi(r_{ij})$ fixed, the anisotropy coefficients $C_1$ and $C_2$ fully determine 
the strength of the induced local anisotropies.
Changing these coefficients consequently broadens or narrows the SO channel, 
thereby controlling the rate of the flow of angular momentum in and out of the spin subsystem.
To study this effect, we repeat our familiar 
thermalization procedure under varying anistoropy coefficients. 
Fig.~\ref{fig:anis_strength} (a) shows the results for varying $C_1$ while $C_2$ set to zero,
whereas in Fig.~\ref{fig:anis_strength} (b), $C_2$ is varied while $C_1$ held constant at $0.2$\;eV. 
As either of the coefficients is increased, we observe a systematic increase in the spin relaxation rate,
which subsequently allows the spin subsystem to reach equilibrium faster.
The stability of our model over such a range of coefficients
promotes its applicability to a wide class of systems with varying SO coupling strengths.
If the interest lies in obtaining realistic relaxation times for the material under investigation,
one can tune $C_1$ and $C_2$ appropriately
in accordance with the spin relaxation data obtained through pump-probe experiments~\cite{Buschhorn2011}. 

So far, our discussion on induced anistoropies was centered on
lattice vibrations as the primary source of symmetry breaking in the local environment. 
Another source of symmetry breaking that commonly occurs in real crystals is the presence of crystallographic defects.
Due to the distortions in the crystal structure surrounding the defect,
SO interactions associated with the nearby atoms will be enhanced significantly~\cite{Defects1, Defects2}.
As a result, the occurrences of defects in the crystal may have a noticeable impact on the overall spin-lattice relaxation.
To investigate this phenomenon, we introduce vacancies into the bcc lattice and observe the relaxation of the spins
as the system is thermalized via a phonon heat bath.
Fig.~\ref{fig:vacancies} shows the time evolution of the spin temperature under varying vacancy concentrations.
As expected, the relaxation rate of the spin subsystem increases as the vacancy concentration is increased.
Site defects could be significantly affected by anisotropic exchange~\cite{dm1,dm2}; this will be studied in future work.

In conclusion, we have developed a generic, phenomenological model for incorporating spin-orbit interactions 
into the simulations of coupled spin-lattice systems.
These interactions are modeled in terms of the local magnetic anisotropies that arise as 
the symmetry of the local crystal structure is broken due to phonons or crystallographic defects.
Our improved approach overcomes the major shortcoming of the original method;
namely, the inability to capture the angular momentum exchange between the lattice and the spin subsystems.
This extends the applicability of the MD-SD approach to the realistic modeling of non-equilibrium processes in magnetic metals and alloys, 
which will, in turn, further our understanding of the microscopic mechanisms of defect evolution, energy dissipation, magnetization dynamics etc.\\[3pt]

\begin{acknowledgments}
D.P.L. thanks the MAINZ (MAterials science IN mainZ) Graduiertenschule for its hospitality. 
This work was sponsored by the “Center for Defect Physics, an Energy Frontier Research Center of the Office of Basic Energy Sciences (BES), U.S. Department of Energy (DOE); 
the later stages of the work of G.M.S. and M.E. was supported by the Materials Sciences and Engineering Division of BES, US-DOE. 
This research used resources of the Oak Ridge Leadership Computing Facility, 
which is supported by the Office of Science of the U.S. Department of Energy under Contract No. DEAC05-00OR22725.
We also acknowledge the computational resources provided by the Georgia Advanced Computing Resource Center.
\end{acknowledgments}

\end{document}